\documentclass[aps,prl,10pt,twocolumn,groupedaddress]{revtex4-1}

\usepackage{graphicx}
\usepackage{hyperref}
\usepackage{amsmath}
\usepackage{amsfonts}
\usepackage{amssymb}
\usepackage{subfigure}
\usepackage{times, color}

\begin{document}

\newcommand{\fixme}[1]{{\bf **FIXME**}\textsc{#1}}

\newcommand{\vgate}{\ensuremath{V_{\rm BG}}}
\newcommand{\expect}[1]{\ensuremath{\langle #1 \rangle}}
\newcommand{\bip}{\ensuremath{B_{\rm IP}}}
\newcommand{\btotal}{\ensuremath{B_{\rm tot}}}
\newcommand{\fmag}{\ensuremath{C}} 

\newcommand{\subfix}{\protect\vphantom{TLSgt} }
\newcommand{\tmag}{\ensuremath{\tau_{\subfix\rm mag}}} %
\newcommand{\tslow}{\ensuremath{\tau_{\subfix\rm slow}}} %
\newcommand{\tfast}{\ensuremath{\tau_{\subfix\rm fast}}} %
\newcommand{\tee}{\ensuremath{\tau_{\subfix\rm e-e}}} %
\newcommand{\trmf}{\ensuremath{\tau_\parallel}} %
\newcommand{\tinel}{\ensuremath{\tau_{\subfix\rm in}}} %
\newcommand{\tsat}{\ensuremath{\tau_{\subfix\rm UCF}^{-1}}} %
\newcommand{\ttrs}{\ensuremath{\tau_{\subfix\rm TRS}}} %
\newcommand{\tip}{\ensuremath{\tau_{\subfix\rm UCF}}} %
\newcommand{\tcurv}{\ensuremath{\tau_{\subfix\rm WL}}} %

\title{Defect-mediated spin relaxation and dephasing in graphene}

\author{M. B. Lundeberg}
\affiliation{Department of Physics and Astronomy, University of British Columbia, Vancouver, British Columbia, V6T1Z4, Canada}
\author{R. Yang}
\affiliation{Department of Physics and Astronomy, University of British Columbia, Vancouver, British Columbia, V6T1Z4, Canada}
\author{J. Renard}
\affiliation{Department of Physics and Astronomy, University of British Columbia, Vancouver, British Columbia, V6T1Z4, Canada}
\author{J. A. Folk}
\email{jfolk@physics.ubc.ca}
\affiliation{Department of Physics and Astronomy, University of British Columbia, Vancouver, British Columbia, V6T1Z4, Canada}

\date{\today}

\begin{abstract}

A principal motivation to develop graphene for future devices has been its promise for quantum spintronics.  Hyperfine and spin-orbit interactions are expected to be negligible in single-layer graphene.  Spin transport experiments, on the other hand, show that graphene's spin relaxation is orders of magnitude faster than predicted.  We present a quantum interference measurement that disentangles sources of magnetic and non-magnetic decoherence in graphene.  Magnetic defects are shown to be the primary cause of spin relaxation, while spin-orbit interaction is undetectably small.  

\end{abstract}

\pacs{73.23.-b}
\maketitle

Spin lifetimes in graphene are remarkable in that they are unremarkable.  The first measurements of electron spin relaxation found 100 picosecond lifetimes\cite{Tombros2007}--similar to what one might expect for conventional metals or semiconductors--and more recent measurements confirm that initial result\cite{Han2010,Han2011}.  These experimental values can be contrasted with much more favourable theoretical predictions: micro- or even milliseconds are expected\cite{Ertler}, due to carbon's low atomic number (weak spin-orbit interaction) and the lack of nuclear spin in the predominant isotope (weak hyperfine interaction). These material properties make graphene an extremely promising material for classical or quantum spintronics\cite{Trauzettel}, but potential applications await an understanding of the practical mechanisms of spin relaxation in the material.

The orders-of-magnitude disconnect between spin relaxation measurements and theory remains one of the most important puzzles in graphene research.  Recent theoretical work has focused on finding a mechanism that would give rise to unexpectedly strong spin-orbit interactions, perhaps associated with random electric fields from the substrate or localized electric fields near adatoms or vacancies.\cite{Ertler}  Another possibility is that the conduction electrons in graphene are strongly coupled to magnetic moments, but this explanation lacks a microscopic picture of where these moments arise.  It is possible to {\em create}  defects with a magnetic moment intentionally in graphene, for example by ion bombardment,\cite{geimimpurity,graphenewlfluorine} but it remains an open question whether paramagnetic defects that may be present in intrinsic (natural) graphene explain the dephasing rate that is observed at low temperatures.

Quantum interference is a powerful tool for studying charge and spin interactions of conduction electrons with their material host.  Random lattice strains, trigonal warping, and atomic-scale disorder in graphene dephase the valley symmetry, and dynamic electron-electron interactions  give rise to inelastic charge dephasing.\cite{graphenewl,tikhonenko,grapheneucf,grapheneucf2,kisat,transitionwl}  Recently it has become clear that additional interactions lead to an apparent saturation of dephasing at temperatures below a kelvin.\cite{kisat,kozikov}  The precise mechanism has yet to be determined, but two of the
proposed mechanisms have a direct bearing on the mystery of graphene's fast spin relaxation: uni-axial spin-orbit interactions,\cite{mccannzz} or defects with a magnetic moment\cite{kozikov, birgeFull}.

This paper presents a quantum interference measurement that distinguishes magnetic and non-magnetic dephasing mechanisms in graphene for the first time. 
We demonstrate that magnetic moments are indeed a significant source of orbital dephasing in graphene, and are the dominant mechanism for spin relaxation\cite{Tombros2007, Han2010, Han2011}; in addition, we find  a non-magnetic dephasing mechanism of similar strength, whose microscopic origin is yet to be determined.

\begin{figure}
\includegraphics{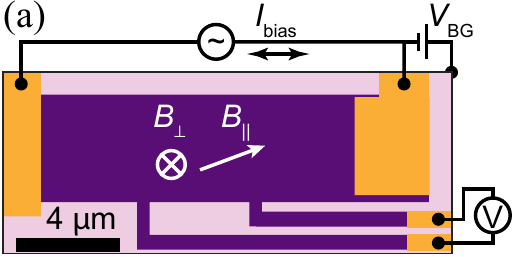}\includegraphics{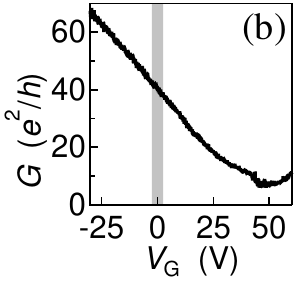}\\
\includegraphics{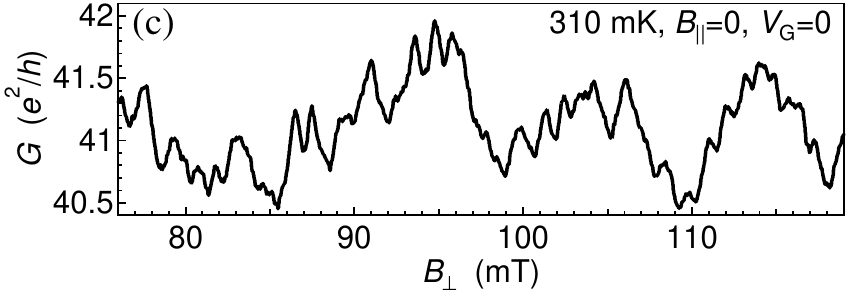}\\
\includegraphics{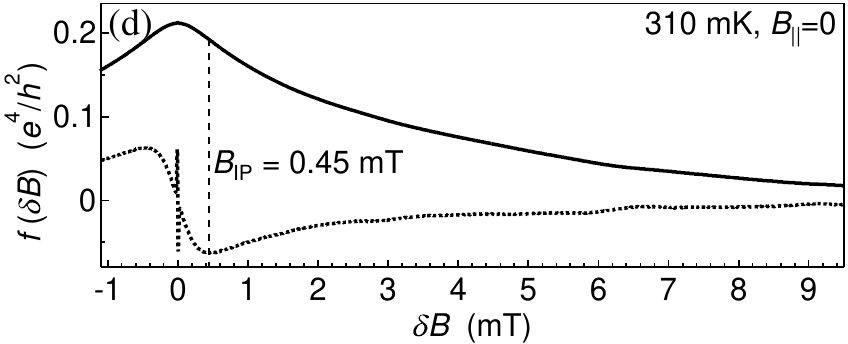}
\caption{\label{fig-intro} (a) Simplified device geometry and measurement setup (see supplement for device and measurement details). Purple regions are graphene, gold indicates leads, and light-shaded regions have been etched away. (b) Conductance as a function of gate voltage; the shaded region indicates the studied interval. (c) A typical UCF trace in perpendicular field. (d) A typical autocorrelation function (solid curve) and its derivative (dotted curve, no vertical scale plotted). The vertical spike in the dotted curve corresponds to the noise peak in the autocorrelation.  The dashed line indicates the inflection point, used as a measure of coherence through Eq.~\eqref{tauucf}.}
\end{figure}

Parameters associated with quantum interference can be extracted from a transport measurement in two ways. Universal conductance fluctuations (UCF) are seen when the phase coherence length is comparable to device dimensions (Fig.~\ref{fig-intro}(a,b,c)), and their statistics may be compared to theoretical predictions. Alternatively, the average low-field magnetoconductance may be fit to weak localization (WL) theory.
From an experimental point of view WL measurements are an easier way to extract a dephasing rate, and historically much more common, but it is difficult or impossible to distinguish different dephasing mechanisms solely from WL.  
This experiment combines UCF and WL measurements to separate and quantify the various sources of dephasing in graphene. We begin by discussing UCF.

A monolayer graphene device was exfoliated onto an SiO$_2$/Si wafer (Fig.~\ref{fig-intro}(a)) and measured in a dilution refrigerator with a two-axis magnet.
Conductance $G$ was measured in a four-terminal configuration, as a function of gate voltage $\vgate$, in-plane and out-of-plane magnetic fields $B_\parallel$ and $B_\perp$, and temperature $T$ (Fig.~\ref{fig-intro}(b,c)).  Similar results were found in a second cooldown (different cryostat) of the same device  after an annealing step.
Conductance fluctuation data  for a narrow range of $\vgate$ were analyzed by their autocorrelation in perpendicular magnetic field, $f(\delta B)$ (Fig.~\ref{fig-intro}(d)).

UCF are dephased by any degrees of freedom in a conduction electron's environment that change faster than the measurement bandwidth (hertz). Such dephasing can be characterized by a rate $\tip^{-1}$ that is the sum of conduction electron scattering rates from uncontrollable dynamic sources.\cite{mpep}
At low temperatures, the dominant contributions to $\tip^{-1}$ are other conduction electrons and dynamic defects in the device, whether magnetic or non-magnetic.
From an experimental point of view, the inflection point of $f(\delta B)$ provides a robust metric of
this rate\cite{inflectionpoint}:
\begin{equation}
\tip^{-1} \approx \frac{2 e D \bip}{3\hbar} ,~{\rm where}\ {}
\frac{\mathrm d^2 f}{\mathrm d\delta B^2} \Big|_{\delta B = \bip} = 0,
\label{tauucf}
\end{equation}
where $D=0.03~\mathrm{m^2/s}$ is the diffusion constant that was calculated from $G$.

\begin{figure}
\includegraphics{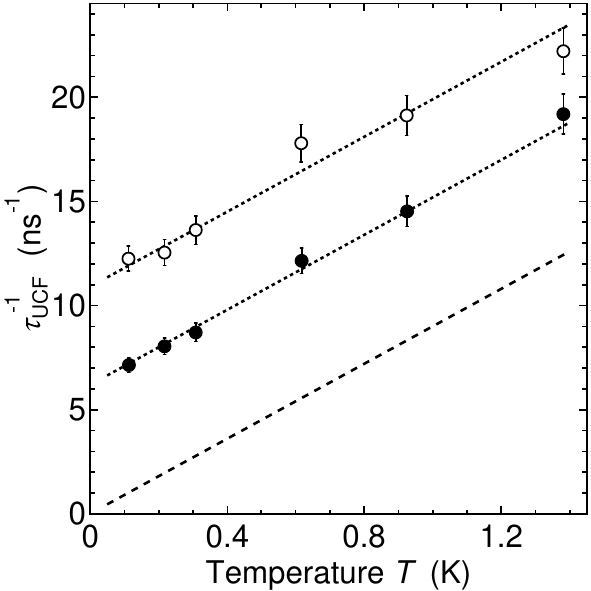}
\caption{\label{fig-tempdep}Dependence of the rate $\tip^{-1}$ on temperature for $B_\parallel = 0$ ($\circ$) and $B_\parallel = 6~\mathrm T$ ($\bullet$).
Dotted lines have slope $9.0~\mathrm{ns^{-1}/K}$ and offsets $\tsat(T{=}0) = 10.9~\mathrm{ns^{-1}}$ (upper line), $\tsat(T{=}0) = 6.2~\mathrm{ns^{-1}}$ (middle line).  The lower, dashed line shows how $\tip^{-1}$ would appear without saturation ($\tsat(T{=}0) = 0$).}
\end{figure}

As seen in Fig.~2, the temperature dependence of $\tip^{-1}$ is linear over the range $T=0.1\cdots 1.4~\mathrm{K}$
with a slope that is close to the value predicted for electron-electron interactions\cite{eeucf,tikhonenko}, and an extrapolated offset  $\tsat(T{=}0)$  that implies a low-temperature saturation of the dephasing rate (the corresponding  length  $\sqrt{D\tip(T{=}0)}=1.7\mu$m is much smaller than the flake dimensions [Fig.~\ref{fig-intro}(a)]).
Whereas certain symmetry-breaking static environments (e.g.~out-of-plane spin-orbit coupling) can induce saturation in WL,\cite{mccannzz,kozikov} they do not lead to a saturation in UCF.\cite{mpep,inflectionpoint}
Instead, the finite $\tsat(T{=}0)$ observed in this experiment indicates the presence of dynamic degenerate defects, that is, defects that do not freeze into a single state as temperature is decreased.
The remainder of this work probes the nature of these degenerate defects: are they magnetic, and how strongly do they interact with the conduction electrons (how fast do they change state)?

Performing the UCF measurement with an in-plane magnetic field shows clearly that some of the defects are magnetic: dephasing is reduced at $B_\parallel = 6~\mathrm{T}$ (Fig.~\ref{fig-tempdep}), indicating that the magnetic moments have been polarized to a static configuration and no longer contribute to $\tip^{-1}$.
Quantitatively, the net change in $\tip^{-1}$ with large $B_\parallel$ is the magnetic scattering rate,  $\tmag^{-1}\approx 4.7\pm 0.5$ ns$^{-1}$.
This rate is, in itself, an important finding, as it corresponds to the spin-flip rate for conduction electrons due to unpolarized magnetic defects.\cite{mpep} 
The data in Fig.~\ref{fig-tempdep} thus prove that magnetic defects induce sufficient spin relaxation to explain previous spin transport measurements in monolayer graphene.\cite{Tombros2007, Han2010, Han2011}

\begin{figure}
\includegraphics{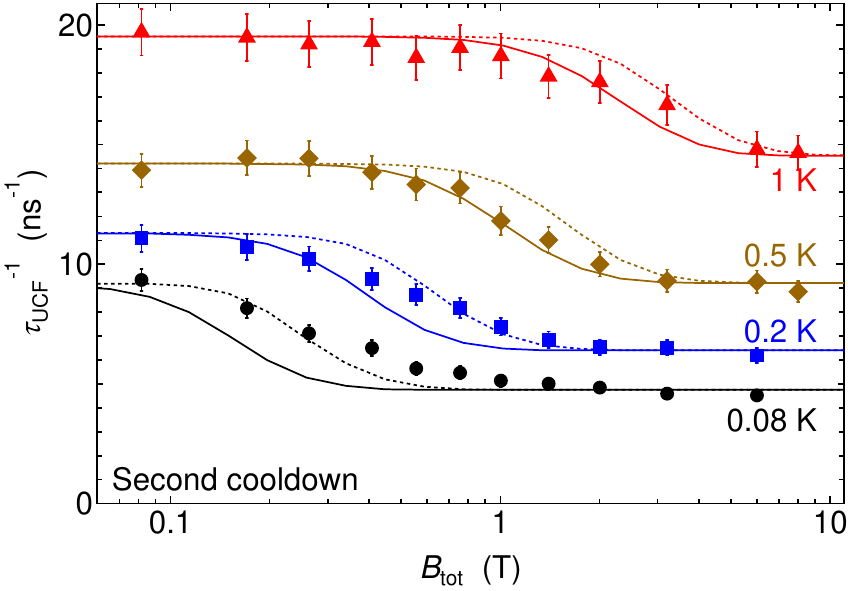}
\caption{\label{fig-fielddep}
Dependence of $\tip^{-1}$ on rms total magnetic field $\btotal = (B_\parallel^2 + \overline{B_\perp^2})^{1/2}$, for various $T$.
Curves show the theoretical crossover for a scattering rate $\tmag^{-1}= 5~\mathrm{ns^{-1}}$ from spin-$\frac{1}{2}$ magnetic defects with $g=2$ (solid), or $g=1$ (dotted).
}
\end{figure}

The crossover to full defect polarization
was analyzed in a second cooldown of this device (Fig.~\ref{fig-fielddep}).  The field required to turn off the dephasing grows with temperature, as expected for the thermodynamics of free magnetic moments.
We obtain the theoretical curves in Fig.~\ref{fig-fielddep} by applying the definition of $\tip^{-1}$ in Eq.~\eqref{tauucf} to numerically simulated UCF with spin-$\frac{1}{2}$ defects (see supplement). At high temperatures the behaviour is consistent with defects of $g=2$.
A significant departure is seen at 200 mK and below, indicating the need for a more careful theoretical treatment of the magnetic defects as quantum objects.\cite{vavilov,micklitz}

The fact that $\tsat(T{=}0)$ does not go to zero at high field indicates an additional saturation mechanism that is apparently non-magnetic, with dephasing rate $\tsat(T{=}0,B_\parallel{=}6~\mathrm T)\approx 6(4)~\mathrm{ns^{-1}}$ [Figs.~\ref{fig-tempdep}(\ref{fig-fielddep}) for the first(second) cooldown].
The data presented so far do not allow us to say more about this non-magnetic mechanism.  Is it merely device noise that limits UCF? Is it a more fundamental inelastic mechanism, such as the two-channel Kondo dephasing that was predicted for metals a decade ago?\cite{tls2}  Along the same lines, it is difficult to ascertain from the UCF data whether the magnetic dephasing results from a Kondo-type interaction of a few defect spins strongly coupled to the electron gas, or from a large number of slowly fluctuating magnetic moments. 

To address these questions we compare the UCF results to an analogous measurement based on WL, which is sensitive to time reversal symmetry (TRS). Like UCF, WL may be dephased by a dynamic environment, but only when the fluctuations occur faster than the dephasing timescale---a cutoff time nine orders of magnitude shorter than the analogous timescale for UCF.
Unlike UCF, WL is also dephased by a static environment if it does not preserve time-reversal symmetry; examples of this are magnetic fields, or spin-flip processes from unpolarized magnetic moments.

\begin{figure}
\includegraphics{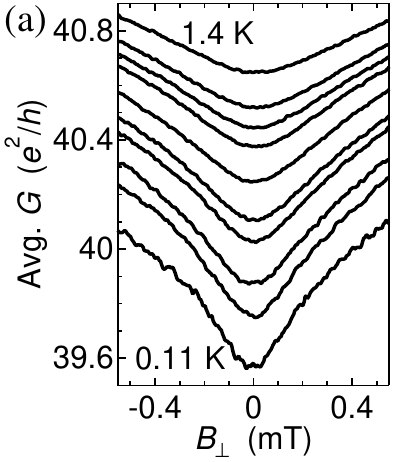}\includegraphics{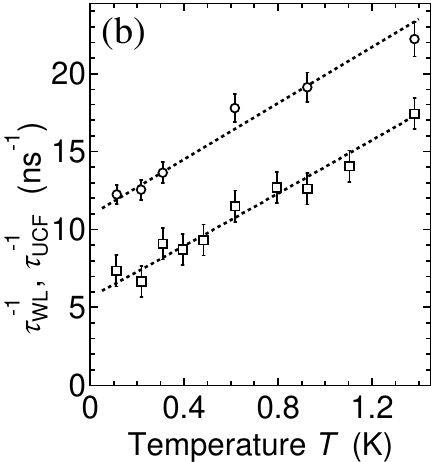}\\
\includegraphics{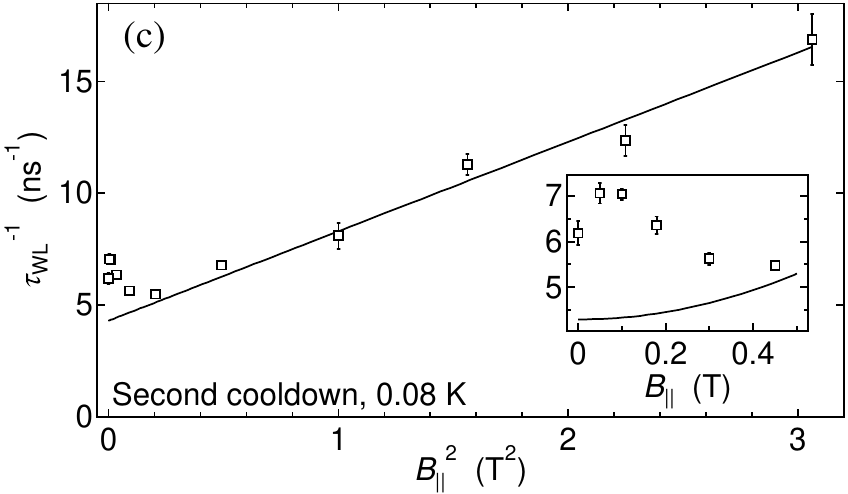}
\caption{\label{fig-wl}
(a) Conductance, averaged over the same $\vgate$ range as used in Fig.~2. (b) Comparison of characteristic rates $\tcurv^{-1}$ [square, extracted from (a)] and $\tip^{-1}$ ($\circ$, from Fig.~\ref{fig-tempdep}) at $B_\parallel = 0$.
The dotted lines are WL and UCF rates expected for electron-electron parameter $9.0~\mathrm{ns^{-1}/K}$, collision rate $\tmag^{-1} = 4.7~\mathrm{ns^{-1}}$ with slow magnetic defects, and non-magnetic offsets of $3.5~\mathrm{ns^{-1}}$ (WL), $6.2~\mathrm{ns^{-1}}$ (UCF).
(c) Dependence of $\tcurv^{-1}$ on $B_\parallel^2$ (same $\vgate$ range as used in Fig.~\ref{fig-fielddep}).
The solid line is $\ttrs^{-1}(B_\parallel) = 4.3~\mathrm{ns^{-1}} + (4.0~\mathrm{ns^{-1}/T^2}) B_\parallel^2$, a fit to the $B_\parallel \geq 1~\mathrm T$ data.
Inset: The same data at low field, plotted linearly in $B_\parallel$.
}
\end{figure}

Graphene's magnetoconductance (Fig.~\ref{fig-wl}(a)) is typically fit to a WL theory\cite{graphenewl} that includes only non-magnetic dephasing mechanisms,
but the UCF data demonstrate that graphene also suffers from significant magnetic dephasing. If the magnetic defects vary slowly, they distinguish the spin-singlet and -triplet channels of the WL correction,\cite{hikamiwl,bobkov,mpep} which complicates the fitting.
The dephasing can be more reliably characterized by extracting the zero-field magnetoconductance curvature to obtain a single rate $\tcurv^{-1}$, defined as
\begin{equation}
\label{taucurv-definition}
      \tcurv^{-1}  \equiv \frac{eD}{\hbar} \left(\frac{3\pi}{4} \frac{h}{e^2} \frac{L}{W} \frac{\mathrm d^2  \overline{G}}{\mathrm d B_\perp^2}\Big|_{B_\perp=0}\right)^{-\frac 12},
\end{equation}
where $\frac{L}{W} = 1.05$ is the device aspect ratio and $\overline G$ is the average conductance.
For slow unpolarized magnetic defects ($B = 0$),\cite{hikamiwl} one expects
\begin{equation}
\label{taucurv-expected}
\tcurv^{-1} \approx \Big(
       \tfrac{3}{2}\big(   \ttrs^{-1}+\tfrac{2}{3}\tmag^{-1}  \big)^{-2}
      - \tfrac{1}{2}\big(    \ttrs^{-1}+2\tmag^{-1}   \big)^{-2}
      \Big)^{-\frac 12}
\end{equation}
where $\tmag^{-1}$ is defined the same as for UCF and $\ttrs^{-1}$ is the summed dephasing rate from other scattering mechanisms that break time reversal symmetry. For fast magnetic defects, on the other hand, $\tcurv^{-1}$ is simply the sum of rates $\approx \ttrs^{-1} + \tmag^{-1}$.\cite{bobkov}

As in the case of UCF, $\tcurv^{-1}$ is seen to increase with temperature, and a zero-temperature offset is clearly observed (Fig.~\ref{fig-wl}(b)).
The common slope with respect to~$T$ reflects the equal effect of electron-electron interactions on WL and UCF\cite{eeucf}, thereby confirming that Eqs.~\eqref{tauucf} and  \eqref{taucurv-definition} are not miscalibrated.   The effect of an in-plane field on WL  (Fig.~\ref{fig-wl}(c)) is more complicated than the analogous measurement for UCF (Fig.~\ref{fig-fielddep}) due to graphene's ripples, which convert the uniform in-plane field to a random vector potential.  This breaks time reversal symmetry,\cite{ripples, mathurbaranger} giving $\ttrs^{-1}(B_\parallel) = \ttrs^{-1}(0) + \beta B_\parallel^2$,
where $\ttrs^{-1}(0)$ is the inelastic dephasing rate from non-magnetic sources and $\beta$ describes the ripple geometry.
In the data $\tcurv^{-1}$ increases sharply from 0 to 50~mT, which can be explained by  the suppression of two WL channels by Zeeman splitting and the resultant transition from Eq.~\eqref{taucurv-expected} to $\tcurv^{-1} = \ttrs^{-1} + \frac{2}{3} \tmag^{-1}$.\cite{vavilov}  Above 50~mT, $\tcurv^{-1}$ decreases at first as the magnetic defects polarize and their dephasing effect vanishes. 
For much higher fields ($B_\parallel > 0.5~\mathrm T$) the defects are fully polarized and $\tcurv^{-1}$ has collapsed to $\ttrs^{-1}$, giving the $B_\parallel^2$ dependence seen Fig.~\ref{fig-wl}(c).

Taking UCF and WL data together, we can draw several conclusions about the mechanisms of spin relaxation and low temperature dephasing in graphene:

1. Scattering from magnetic defects induces a spin flip rate $\tmag^{-1}=5~\mathrm{ns^{-1}}$.  This is seen directly as the field-induced suppression of $\tip^{-1}$ (Fig.~\ref{fig-fielddep}). The smaller field-induced suppression of $\tcurv^{-1}$  ($\sim 2$ ns$^{-1}$, Fig.~\ref{fig-wl}(c)) is consistent with the weaker contribution of $\tmag^{-1}$ to $\tcurv^{-1}$ for slow magnetic defects, i.e., those that change slowly on the dephasing timescale but  fast enough to dephase UCF.

2. Spin-orbit interactions can be excluded as a significant contribution to spin relaxation in graphene.   Spin-orbit coupling would generate either antilocalization at $B_\parallel = 0$ or a much larger decrease in $\tcurv^{-1}$ for small $B_\parallel$,  depending on the spin-orbit symmetry\cite{mccannzz}; moreover the temperature dependence $\tip^{-1}(T)$ would be nonlinear\cite{inflectionpoint}.  None of these effects are observed. 

3. WL and UCF each indicate a non-magnetic component to the saturation in dephasing. For the second cooldown, the non-magnetic rate for WL dephasing was $\ttrs^{-1}(T{=}0) \approx 3.5 \pm 1$ ns$^{-1}$ after subtracting the contribution from electron-electron interactions, identical to the corresponding rate for UCF, $\tsat(T{=}0, B_\parallel{=}6~\mathrm T) \approx 3.8\pm 0.2$ ns$^{-1}$.  This indicates that the non-magnetic source breaks time-reversal symmetry, and must therefore change rapidly on the dephasing timescale.  For the first cooldown, the UCF rate was higher ($\tsat(T{=}0, B_\parallel{=}6~\mathrm T) \approx 6.2 \pm 0.3 $ ns$^{-1}$) while the WL rate was $\ttrs^{-1}(T{=}0) \approx 3.5 \pm 1$ ns$^{-1}$; the higher UCF rate may be attributed to  
defects with dynamics too slow to break time reversal symmetry.

Both magnetic and non-magnetic dephasing mechanisms limit coherence in graphene below 1K.
The magnetic scattering rate is too large to be explained by remote magnetic moments, requiring instead that the magnetic defects are electronically coupled to the graphene.
Recent WL data\cite{kozikov} suggests that the magnetic defects may be midgap states at the Dirac point, formed at vacancies or edges.
For the non-magnetic dephasing, we can rule out bistable charge systems in the SiO$_2$ substrate; their broadly-distributed level splittings would produce a rate proportional to $T$.\cite{tls2} The data instead suggest a class of nearly degenerate non-magnetic defects in the graphene itself, whose microscopic origin is yet to be determined.

\begin{acknowledgments}
We acknowledge helpful discussions with V.~I.~Falko. Graphite crystals were provided by D.~Cobden. J.~R.~acknowledges funding from the CIFAR JF Academy and from the Max Planck-UBC Center for Quantum Materials. This work was funded by CIFAR, CFI, and NSERC.\end{acknowledgments}

\newcommand{\tdyn}{\ensuremath{\tau_{\subfix\phi}}}

\newcommand{\topp}{\ensuremath{\tau_{\uparrow\downarrow}}}
\newcommand{\tsame}{\ensuremath{\tau_{\uparrow\uparrow}}}
\newcommand{\tsym}{\ensuremath{\tau_{\rm sym}}}

\newcommand{\tinf}{\ensuremath{\tau_\infty}}
\newcommand{\tpol}{\ensuremath{\tau_{\rm mag}}}

\appendix*
\clearpage

\numberwithin{figure}{section} %
\setcounter{figure}{0} %
\renewcommand{\thefigure}{S:\arabic{figure}} %

\numberwithin{table}{section}
\setcounter{table}{0}
\renewcommand{\thetable}{S:\Roman{table}}

\renewcommand{\theequation}{S\arabic{equation}} %

\renewcommand{\textfraction}{0.1}
\renewcommand{\topfraction}{0.8}
\renewcommand{\bottomfraction}{0.8}

\begin{center}
\section*{Supplementary information}
M.~B.~Lundeberg, R.~Yang, J.~Renard, J.~A.~Folk

 \today
\end{center}

This supplement covers auxiliary topics relevant to the primary text titled {\em Defect-mediated spin relaxation and dephasing in graphene}. The figures in this supplement should be viewed in colour. These topics are covered in the following sections:
\begin{enumerate}
\item The sample fabrication details are described, as well as the measurement and analysis methods.
\item The expected influences on the device's temperature are discussed, in particular the overheating caused by the applied bias current. Two measured effects show the ability to reach low temperatures: the electron-electron interaction correction to conductivity, and the energy correlation width of UCF.
\item We discuss the accuracy of the UCF autocorrelation inflection point when used as a measure of the decoherence rate.
\item We describe the classical magnetic defect polarization model of UCF, which was used to interpret the crossover of $\tip^{-1}$ in $B_\parallel$.
\end{enumerate}

\subsection{Materials and methods}

\begin{figure}[t]
\includegraphics{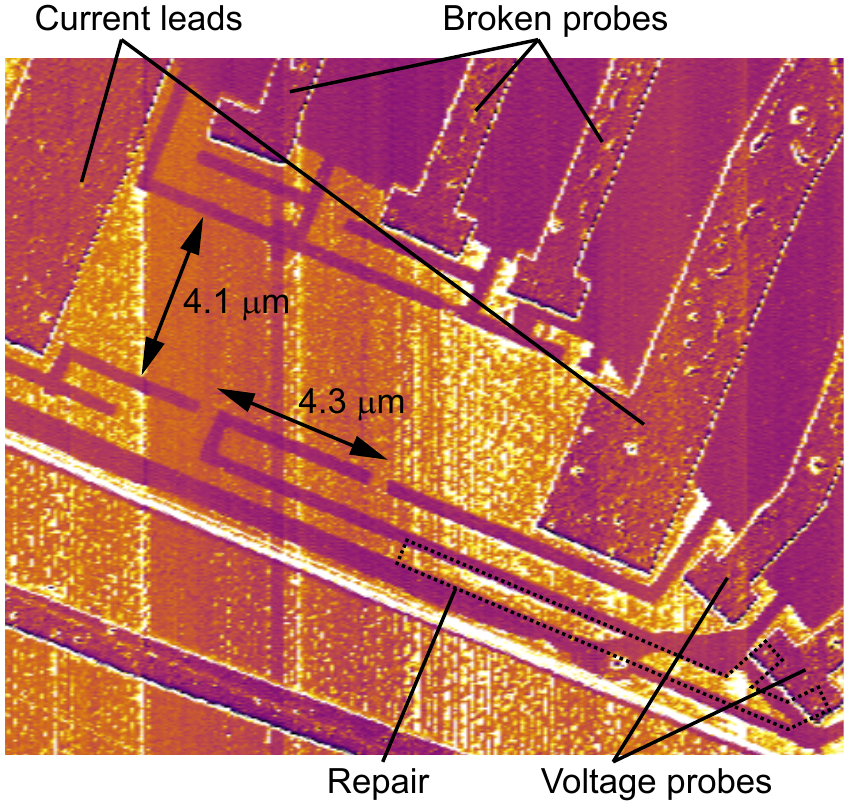}
\caption{\label{supp-afm}Phase image from atomic force microscopy, taken after deposition of gold leads, plasma etching, and a preliminary measurement. Electrostatic shock damage is visible as gaps in the graphene leads. After this image was taken, the dashed region was filled with gold, repairing this voltage probe and allowing the primary measurements to take place. The channel width and the spacing between voltage probes are indicated.}
\end{figure}

Single-layer graphene was deposited onto an SiO$_2$/Si wafer by exfoliation from a natural graphite crystal, then contacted by CrAu metallization and plasma etched into a channel.  Figure~\ref{supp-afm} shows the device as imaged by a scanning probe microscope.
The device was designed for four-probe measurements, with a channel large enough to be within the quasi-2D regime of phase coherence, yet small enough that the conductance fluctuations were not self-averaged. In order to avoid non-local coherent effects from the metals, the voltage probes were etched into long strips of graphene (longer than the coherence length). Note that some damage to the voltage probes occurred before the measurements described in the main text, so a second CrAu deposition step was necessary.

After fabrication the device was annealed for 30 minutes at $400~\mathrm{^\circ C}$ in N$_2$:H$_2$.
The device was measured in dilution refrigerators equipped with two-axis magnets, enabling independent control of magnetic field components in the plane of the graphene ($B_\parallel$) and perpendicular to it ($B_\perp$).
Differential conductance $G$ was measured by standard four-terminal lock-in techniques at 870 Hz (Primary Fig.~1A).
The carrier density was controlled by the back gate voltage $\vgate$ through 275~nm of SiO$_2$, giving a typical conductance curve with the minimum near $\vgate \sim 50~\mathrm{V}$ (Primary Fig.~1B).
In the first cooldown we examined a narrow range of gate voltage $\vgate = -2.5 \cdots 2.5 ~\mathrm{V}$ ($\vgate = -1 \cdots 1 ~\mathrm{V}$ for the $B_\parallel$-dependence data), where the carrier density was $n_s \approx -4 \times 10^{12}~\mathrm{cm^{-2}}$ and the diffusion constant was $D \approx 0.030~\mathrm{m^2/s}$ (estimated from the conductivity and density of states, via the Einstein relation).
A second cooldown of the same device, after a second annealing procedure in N$_2$:H$_2$, produced similar results, with parameters $\vgate = -5\cdots5~\mathrm V$, $n_s \approx -3 \times 10^{12}~\mathrm{cm^{-2}}$, $D \approx 0.033~\mathrm{m^2/s}$. The data in the main text and this supplement are from the first cooldown, except where otherwise noted.

Conductance fluctuation data were analyzed by their autocorrelation in perpendicular magnetic field, $f(\delta B)$, defined as
\begin{widetext}
\begin{align}
f(\delta B) & = \big\langle \delta G(B_\perp,\vgate)\,\delta G(B_\perp+\delta B,\vgate) \big\rangle_{B_\perp,\vgate} \\
& = \frac{1}{[\sum_{\vgate}]} \sum_{\vgate} \left[ \frac{1}{B_2-\delta B - B_1} \int_{B_1}^{B_2-\delta B} \delta G(B_\perp,\vgate) \,\delta G(B_\perp+\delta B,\vgate) \,\mathrm dB_\perp \right]. 
\label{fdef}
\end{align}
\end{widetext}
Here, $\delta G(B_\perp,\vgate)$ is the fluctuating part of conductance, obtained from the raw data $G(B_\perp,\vgate)$ by subtracting off a smooth background function.
The conductance was scanned over a range $B_1 \sim 50~\mathrm{mT}$ to $B_2 \sim 150~\mathrm{mT}$, for ten to twenty different values of $\vgate$ spread over the narrow range of gate voltage noted above; the averaging over $\vgate$ was done to gather more fluctuations and thereby improve the statistical accuracy of \eqref{fdef}.
The smooth background, used to obtain $\delta G$, was obtained by fitting $G(B_\perp,\vgate)$ to a polynomial of the form
\begin{equation}
g_0 + g_1 B_\perp + g_2 B_\perp^2 + g_3 \vgate + g_4 \vgate^2 + g_5 B_\perp \vgate.
\end{equation}
for free parameters $\{g_i\}$.

\subsection{Device temperature}

There are many claims of saturating physics at low temperatures, and it is always crucial to establish that the device temperature is in fact as low as the temperature of the cryostat. Otherwise it may simply be the temperature, not the physics, that is saturating.
This is of great concern for experiments below about 50 mK where it soon becomes difficult to properly heat-sink the device in the presence of virtual heat leaks, paramagnetic or nuclear heating by a magnetic field, or Joule heating due to measurement.

Regardless of the ease of reaching a 100 mK device temperature, we have taken several precautions and performed tests to remove all doubt about the accuracy of the temperature values. Since we examine only temperatures above $\sim 100$ mK in the primary text, and the saturation is already noticeable above $\sim 500~\mathrm{mK}$, the following discussion is expected to be of supplementary interest. 

\subsubsection{\label{sec:heating}Expected influences on device temperature}

The measurements in the first cooldown were carried out in the same dilution refrigerator and magnet system as our previous experiments of coherent effects in graphene, described in Refs.~\onlinecite{nphys,ripples}. Since those experiments, we have added a new wiring and shielding system to the refrigerator. The same system has been added to a second cryostat, used for the second cooldown.
Heat-sinking of the device is ensured by providing a $10~\mathrm{m\Omega}$ wiring path from the chip carrier to a metal-epoxy-metal heat sink of $\sim 200~\mathrm{mm^2}$ area per wire; this $10~\mathrm{m\Omega}$ path is free of solder (a superconductor), guaranteeing efficient heat sinking at all magnetic fields.
High-frequency signals originating from the warm parts of the refrigerator are removed on every wire by a series of three surface-mount RC low-pass circuits (1.6~MHz, 1.6~MHz, 16~MHz), followed by a 2.5~m long lossy RF transmission line; all filters are thermally anchored to the mixing chamber so they do not add thermal noise. 
The entire arrangement (filters, device, and heat sink) is enclosed by a coaxial radiation shield also attached to the mixing chamber.

In fact, the greatest source of heating in this experiment was intentional: In order to expedite the measurement of conductance fluctuations with low noise levels, we applied nonnegligible bias currents through the graphene.
The bias current induces Joule heating, and due to the finite thermal conductivity of the graphene this overheats the middle of the device, which is the measured region.
We calculate the overheating by assuming the device is thermally a one-dimensional system and that all heat must leave through the wires by electronic heat conduction.
Based on the Wiedemann-Franz law, one obtains the rms temperature at the hottest point (the center of the device), as\cite{birgeFull}
\begin{equation}
T = \sqrt{T_{\rm cryostat}^2 + T_{\rm bias}^2}, \quad T_{\rm bias} = \frac{\sqrt{3}e}{2\pi k} I_{\rm bias}R
\label{biasheat}
\end{equation}
where $R$ is the total electrical resistance from the current source's heat sink to the current drain's heat sink, and $I_{\rm bias}$ is the rms current. For the $\vgate $ range which was the focus of this work, $R \approx 1750~\Omega$.

\begin{figure}
\includegraphics{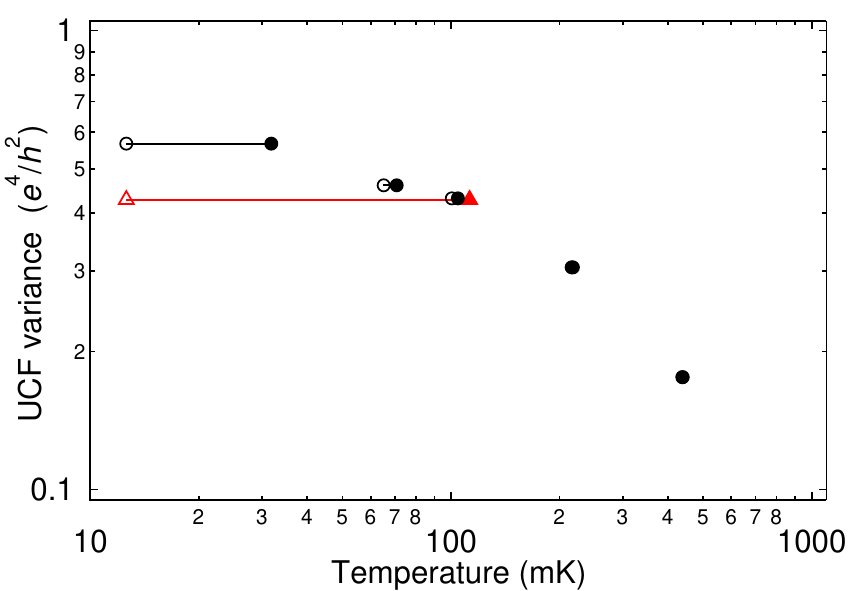}
\caption{\label{supp-varcheck}Bias overheating effect: the variance, computed from $G(\vgate)$ over $\vgate = -5\cdots 5~\mathrm{V}$, as a function of temperature. Here $B_\parallel = 0$, $B_\perp \sim 100~\mathrm{mT}$, and $I_{\rm bias} = 5~\mathrm{nA}$ ($\circ$, $\bullet$) or 20~nA (\textcolor{red}{$\vartriangle$}, \textcolor{red}{$\blacktriangle$}).
Open symbols show the uncorrected $T_{\rm cryostat}$, and filled symbols the corrected temperature $T$ via \eqref{biasheat} with $R = 1750~\mathrm{\Omega}$.
}\end{figure}

Figure~\ref{supp-varcheck} shows the variance of UCF for different bias currents and temperatures. So long as $kT/\hbar$ is not much less than the decoherence rate, the variance is sensitive to temperature\cite{bergmann,mpep} and so the accuracy of \eqref{biasheat} can be easily detected by observing the variance. The figure shows how the temperature $T \approx 110~\mathrm{mK}$ (which has UCF variance $ 0.43~e^4/h^2$) can be produced either by a warm $T_{\rm cryostat} \approx 100~\mathrm{mK}$ or a high bias current $T_{\rm bias} \approx 110~\mathrm{mK}$, as predicted by \eqref{biasheat}. The variance saturates below $\sim 100~\mathrm{mK}$, which is consistent with $kT/\hbar$ falling below the saturated decoherence rate.\cite{inflectionpoint} Subsection \ref{ecorrelations} explores another aspect of this data.

In the first cooldown, equation \eqref{biasheat} had its greatest relevance for the lowest temperatures $T = \{110~\mathrm{mK}, 220~\mathrm{mK}, 310~\mathrm{mK}\}$ of the UCF $B_\perp$-correlation data set discussed in the primary text; these temperatures were achieved by applying $I_{\rm bias} = 20~\mathrm{nA}$ with $T_{\rm cryostat} = \{13~\mathrm{mK}, 190~\mathrm{mK}, 290~\mathrm{mK}\}$. At higher temperatures we used higher currents (up to 45 nA). Similar bias currents were used for the WL measurements (except for the 110 mK data, which used 10 nA). Similar bias currents were used in the second cooldown.

\subsubsection{Test of non-saturating temperature: e-e interaction correction to conductivity}

There are other aspects of the device which demonstrate that the temperature can be reduced appropriately.
One common transport test of low temperatures is to examine the temperature dependence of ensemble-averaged conductivity at nonzero magnetic field. A non-saturating contribution to conductivity, proportional to $\ln T$, is expected from electron-electron interactions:
\begin{equation}
\langle\sigma\rangle = \sigma_0 + \frac{e^2}{\pi h} \left(1 + c\left [1-\frac{\ln(1+F)}{F}\right] \right) \ln T,
\label{sigmaee}
\end{equation}
where $\sigma_0$ is an offset, $F = -0.1$ is predicted for graphene on SiO$_2$, and $c=3$ at the low temperatures we investigate here.\cite{grapheneee,grapheneeej}
This effect should not be confused with the electron-electron decoherence which indirectly causes a temperature-dependent conductance, via the WL effect.
In fact, measurements of \eqref{sigmaee} should be carried out in a nonzero magnetic field to break time reversal symmetry and thus suppress the coherence-sensitive part of WL; this allows a simple comparison with \eqref{sigmaee}.

\begin{figure}
\includegraphics{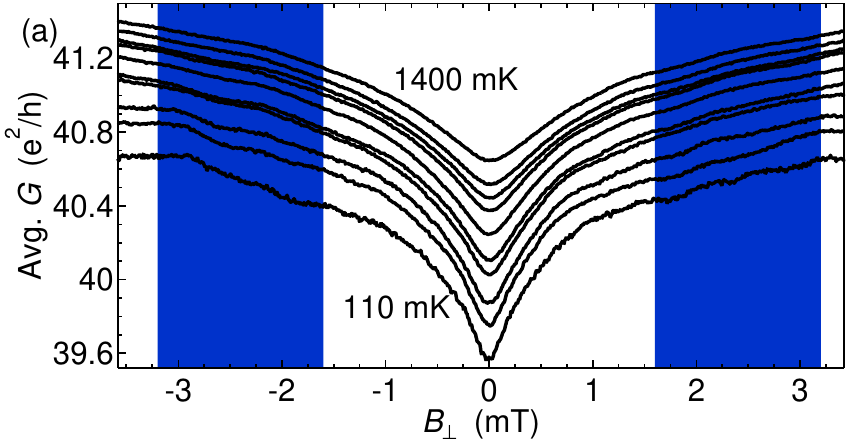}
\includegraphics{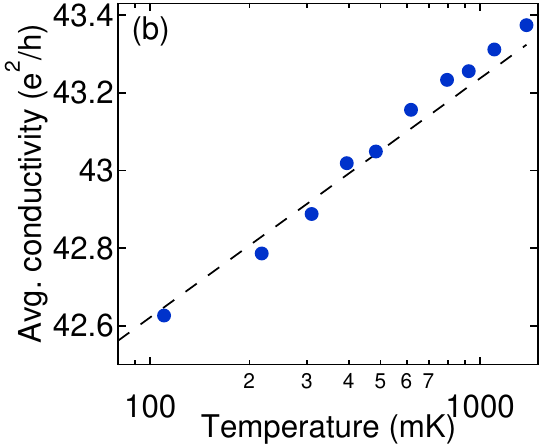}
\caption{\label{supp-tempwl}Electron-electron interaction contribution to conductivity. ({\em a}) $\vgate$-averaged magnetoconductance data expanded from the primary text. Shaded regions indicate the field-averaging regions used for panel b. ({\em b}) Conductivity averaged over both $\vgate$ and $B_\perp$ (circles) plotted against log temperature.
The dashed line shows the slope predicted by theory \eqref{sigmaee}.
}\end{figure}

In Figure \ref{supp-tempwl} we show the conductivity over a small range in $B_\perp$ near zero, averaged over $\vgate$ -- this is a wider range of the WL data displayed in the primary text (from the first cooldown). Above 1~mT the WL contribution to conductivity is insensitive to the coherence and thus is temperature-independent; here, WL merely causes a B-dependence of the offset $\sigma_0$ in Eq.~\eqref{sigmaee}. Thus, by averaging conductivity over a consistent range of magnetic fields above 1 mT, we can test \eqref{sigmaee}. This $\vgate,B_\perp$-averaged conductivity (Fig.~\ref{supp-tempwl}) shows the expected logarithmic temperature dependence and the slope is close to the expected value.

\subsubsection{\label{ecorrelations}Test of non-saturating temperature: UCF thermometry}

As a second test of temperature, we note the recent report\cite{falkoucf} that UCF themselves can provide information on the electron temperature in back-gated devices such as ours. In this case, we measure conductance as a function of back-gate voltage (not magnetic field), and subtract a linear background to yield the conductance fluctuations $\delta g(\vgate)$. Then, $\vgate$ is converted to Fermi energy $E_F$ using the ideal graphene relation $E_F = \hbar v \sqrt{\pi \alpha |\vgate - V_0|}$, where: $v = 10^6~\mathrm{m/s}$ is the Fermi speed, $\alpha = 8\times 10^{10}~\mathrm{cm^{-2}/V}$ is the gate capacitance density, and $V_0 = 45~\mathrm{V}$ is the presumed location of the Dirac point in this device. This gives us the UCF in energy, $\delta g(E_F)$, from which we calculate the autocorrelation function in energy:
\begin{equation}
f_E(\delta E) = \frac{1}{E_2 - E_1}\int_{E_1}^{E_2} \delta g(E_F)\delta g(E_F+\delta E) \,\mathrm dE_F.
\label{fedef}
\end{equation}
When $kT/\hbar$ greatly exceeds the decoherence rate $\tdyn^{-1}$, the correlation function's shape is entirely determined by thermal smearing, so the half-width of $f(\delta E)$ is expected\cite{falkoucf} to be $E_{1/2} = 2.72 kT$, and the inflection point is expected\cite{inflectionpoint} to be $E_{\rm IP} = 2.14 kT$. In fact, at low temperatures this condition is not satisfied: we have $kT/\hbar \sim \tdyn^{-1}$  due to the saturating coherence which is the topic of the primary text, so the decoherence rate also influences the width of $f(\delta E)$. In this situation, $E_{1/2}$ (or $ E_{\rm IP}$) can still be converted to $T$ though in a more complicated manner, by taking into account the effect of $\tdyn^{-1}$ (which is known from $\bip$).\cite{inflectionpoint}

\begin{figure}
\includegraphics{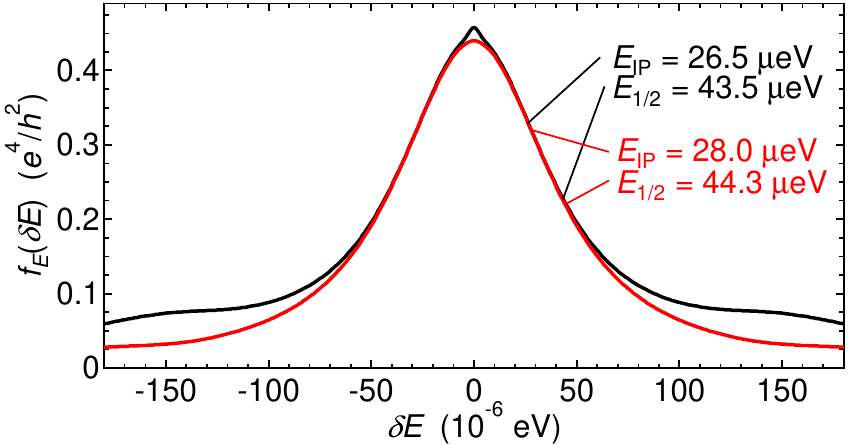}
\caption{\label{supp-ecorrelation2}Energy autocorrelations of UCF for $B_\parallel = 0$, $B_\perp \sim 100~\mathrm{mT}$. The black curve corresponds to $I_{\rm bias} = 5~\mathrm{nA}$ with $T_{\rm cryostat} = 100~\mathrm{mK}$. The red curve corresponds to $I_{\rm bias} = 20~\mathrm{nA}$ with $T_{\rm cryostat} = 13~\mathrm{mK}$. The small peak on top of the black curve is an artifact of measurement noise. Correlations were computed from $G(\vgate)$ measured over $\vgate = -5\cdots5~\mathrm V$. [These are two of the data points in Fig.~\ref{supp-varcheck}.]
}\end{figure}

Figure \ref{supp-ecorrelation2} shows the energy autocorrelation from two conductance traces in the first cooldown at $T\approx 100~\mathrm{mK}$, where $\tdyn^{-1} = 12~\mathrm{ns^{-1}}$. By taking into account the value $\tdyn^{-1} = 12~\mathrm{ns}^{-1}$ that was measured in this situation, we compute the temperature from the energy widths.\cite{inflectionpoint} In one case the temperature is mainly due to $T_{\rm bias}$, and its energy widths correspond to 120~mK (whereas \eqref{biasheat} predicts 113~mK). In the other case the temperature is mainly due to $T_{\rm cryostat}$, and its energy width correponds to 115~mK (whereas \eqref{biasheat} predicts 104~mK). The use of either $E_{1/2}$ or $E_{\rm IP}$ yielded the same temperature to within 4\%, for these correlation functions.

\begin{figure}
\includegraphics{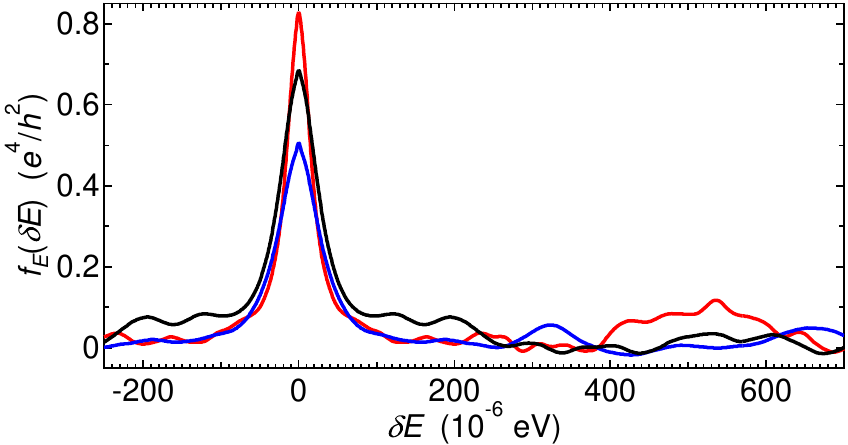}
\caption{\label{supp-ecorrelation}Energy correlations of UCF for $(B_\parallel,B_\perp)$ values of \textcolor{black}{$(0,0)$}, \textcolor{blue}{$(0,50~\mathrm{mT})$}, and \textcolor{red}{$(6~\mathrm{T},50~\mathrm{mT})$}. Correlations were computed from $G(\vgate)$ measured over $\vgate = -5\cdots5~\mathrm V$, with $I_{\rm bias} = 4~\mathrm{nA}$ and $T_{\rm cryostat} \sim 13~\mathrm{mK}$.
}\end{figure}

Figure \ref{supp-ecorrelation} shows autocorrelations taken at even lower temperatures, $T \approx 26~\mathrm{mK}$ according to Eq.~\eqref{biasheat}, at various magnetic fields.
The $E_{1/2}$ correspond to temperatures $T \approx 50~\mathrm{mK}$ for the $B_\parallel = 6~\mathrm{T}$ data (based on observed high-field dephasing $6~\mathrm{ns^{-1}}$), or $T \approx 65~\mathrm{mK}$ for the $B_\parallel = 0$ data (based on observed low-field dephasing $11~\mathrm{ns^{-1}}$). Similar values are obtained from $E_{\rm IP}$.
It is strange that the observed widths of $f(\delta E)$ are too large at this low temperature, however we can be confident that the temperature is not saturating in the 0.1--1.4 K region examined in the primary text.

\subsection{Correspondence between inflection point and decoherence rate}

The rate $\tip^{-1}$ defined from the inflection point $\bip$,
\begin{equation}
\tip^{-1} \equiv 2 e D \bip/3\hbar,
\label{tucfapprox}
\end{equation}
only corresponds to the true UCF decoherence rate $\tdyn^{-1}$ under certain conditions, as explained in Ref.~\onlinecite{inflectionpoint}. We list several conditions below; some conditions are nearly violated and hence there are some small systematic errors in Def.~\eqref{tucfapprox} ({\em error} meaning the difference between $\tdyn^{-1}$ and $\tip^{-1}$).
Some of these errors are known well enough to be corrected, however we have chosen not to correct them in order to retain the simplicity of definition \eqref{tucfapprox}.
In any case, none of these error sources are able to artificially generate the saturation that is the topic of the primary text.

\subsubsection{Limiting case: quasi-2D, thermally-smeared}
Definition~\ref{tucfapprox} assumes the quasi-2D thermally smeared regime, where the coherence length $\sqrt{D\tdyn}$ is smaller than device dimensions, yet is larger than the `thermal length' $\sqrt{D\hbar/kT}$.

The transition from quasi-2D behaviour\cite{inflectionpoint} ($\bip^{\rm 2D} = \frac{3}{2} \hbar/[eD\tdyn]$) to quasi-1D behaviour\cite{inflectionpoint} ($\bip^{\rm 1D} = \sqrt{6} \hbar/[eW\sqrt{D\tdyn}]$ for width $W$) should occur in the vicinity of $\tdyn^{-1} \sim D/W^2$.
In the present device $W = 4.3~\mathrm{\mu m}$, so the crossover would be when  $\tdyn^{-1}$ drops below $1.6~\mathrm{ns^{-1}}$.
The lowest $\tip^{-1}$ measured was $5~\mathrm{ns^{-1}}$, at high $B_\parallel$ and the lowest temperature; this is still a factor of three higher than the crossover value, but it is perhaps possible that $W$ begins to influence $\tip^{-1}$ at this point.
We are unaware of any predictions of behaviour in the crossover and so there is no simple way to quantify this error.
As an unlikely worst case the crossover could be a gradual quadratic crossover ($\bip = \bip^{\rm 2D} \cdot [ 1 + \frac{8}{3}D\tdyn/W^2]^{1/2} $), meaning that all of our extracted $\tip^{-1}$ values would be $2~\mathrm{ns^{-1}}$ higher than the actual $\tdyn^{-1}$.

The thermal smearing limit holds up to\cite{inflectionpoint} $\bip \lesssim kT/eD$. For the lowest temperatures ($\sim 100~\mathrm{mK}$) at zero $B_\parallel$, the values become as large as $\bip = 1.3 kT/eD$; in this situation the error in $\tip$ due to the thermal smearing approximation  is 10\%. For all other data points we have $\bip \leq 0.7 kT/eD$ giving errors less than 4\%.

\subsubsection{Background subtraction biases}
The systematic errors originating from the conductance background estimation are very small for the inflection point, compared to other aspects of the correlation function such as variance or half-width.\cite{inflectionpoint}
Based on the considerations in Ref.~\onlinecite{inflectionpoint}, the resulting bias errors in $\tip^{-1}$ are negligible, $<3\%$ for the high-temperature data and $<1\%$ for the low-temperature data.

\subsubsection{Perturbation by symmetry-breaking disorder}
When there are static symmetry-breaking forms of disorder, some (but not all) modes of UCF may be suppressed. If the symmetry-breaking rate is comparable to the decoherence rate $\tdyn^{-1}$, then the correlation function's shape is distorted\cite{chandrasekhar,inflectionpoint} and Def.~\eqref{tucfapprox} loses some accuracy. The symmetry-breaking rate causes no change if it is much smaller than the decoherence rate; conversely if the rate is very large then the suppressed modes are too weak to contribute.

Graphene's two valley symmetry breaking rates\cite{grapheneucf} were measured by the usual WL fitting formula\cite{graphenewl} at fields above 1 mT. We extract $\tau_i^{-1} = 70\pm 20~\mathrm{ns^{-1}}$ and $\tau_*^{-1} = 4000\pm 200~\mathrm{ns^{-1}}$, much like seen in other studies.\cite{tikhonenko,ripples} These rates are very large compared to $\tip^{-1}$ and hence the valley-related modes have little influence. The influence of the $\tau_i$-suppressed mode on $\tip^{-1}$ can be calculated\cite{inflectionpoint} and is less than 10\% for the high temperature (1.4K) data, and less than less than 4\% for the low temperature data.
The larger rate $\tau_*^{-1}$ gives even smaller effects that matter only above 10 K.

Graphene's spin-orbit rates are expected to be very low $\ll 1~\mathrm{ns^{-1}}$ and as yet have not been experimentally determined. If stronger than expectations, then spin-orbit interactions would dephase some UCF modes and cause some deviations in $\tip^{-1}$. A very large ($\gtrsim 5~\mathrm{ns^{-1}}$) spin-orbit rate would however produce a specific temperature-dependent shift\cite{inflectionpoint} in $\tip^{-1}$ which is not observed in the data.

In conclusion, $\tip^{-1}$ measures the rate $\tdyn^{-1}$ with typical accuracy better than $10\%$ for the $B_\parallel=0$ and  $B_\parallel=6~\mathrm{T}$ data.
There is however another form of symmetry-breaking disorder that is present when $B_\parallel$ is applied, which does lead to significant changes in the crossover of $\tip^{-1}(\btotal)$; this is the topic of the following section.

\subsection{Quasi-2D UCF correlations in an in-plane magnetic field}

A large in-plane magnetic field induces a Zeeman splitting $E_Z = 2\mu_B \btotal$ of conductance fluctuations, while simultaneously the magnetic defects gain an average polarization.\cite{vavilov}
This causes the UCF field correlation function $f(\delta B)$ to evolve in a nontrivial manner as the $\btotal$ is turned on. What determines the inflection point $\tip^{-1}$ in this case?

\subsubsection{\label{classicalmodel}Model}

To answer this question, we simulate the theoretical correlation function of UCF including high-field effects, and apply Def.~\eqref{tucfapprox} to predict the value of $\tip^{-1}$.
This simulation requires us to move beyond the one-parameter correlation of the primary text to the more general two-parameter correlation function of UCF\cite{lsf,bergmann,inflectionpoint}
\begin{equation}
F(\delta B, \delta E) = \overline{\delta G(B_\perp,E) \,\delta G(B_\perp+\delta B, E+\delta E)},
\label{ftwo}
\end{equation}
where $\delta G(B_\perp,E)$ is the fluctuating part of conductance at perpendicular magnetic field $B_\perp$ and energy $E$.
The ordinary one-parameter magnetoconductance correlation, Eq.~1 in the main text, corresponds to a measurement of $f(\delta B) = F(\delta B, 0)$, i.e.~it corresponds to $\delta E = 0$, since we only compute the correlation between conductances taken at the same $\vgate$ (same $E$).

In the analysis of UCF where the spin degeneracy is broken (by e.g.~spin-orbit interactions\cite{chandrasekhar}, frozen magnetic impurities\cite{bobkov}, polarized impurities\cite{vavilov}, or Zeeman splitting\cite{vavilov}), the full $F$ correlation function is written in terms of up to four distinct modes, and each of these modes can be written in terms of the spinless correlation $\tilde F$ which is given by the generic UCF theory.
This spinless correlation function depends on a single dephasing rate $\tau_\phi^{-1}$ and temperature $T$, and is fully specified as $\tilde F[\tau_\phi^{-1},T](\delta B, \delta E)$.
In the procedure below we use the quasi-2D spinless $\tilde F$ computed by the procedure in Ref.~\onlinecite{inflectionpoint} since our device is in the quasi-2D regime. The results can be extended to the quasi-1D case by using the spinless correlation function found in Ref.~\onlinecite{falkoucf} (with the field-dependence noted in Ref.~\onlinecite{beenakker1d}).

For the case of classical magnetic defects and no spin-orbit interaction\cite{vavilov},
\begin{align}
F(\delta B, \delta E) 
&= 2 \tilde F[\tsame^{-1},T](\delta B, \delta E) \nonumber \\
&\quad + \tilde F[\topp^{-1},T](\delta B, \delta E + E_Z) \nonumber \\
&\quad + \tilde F[\topp^{-1},T](\delta B, \delta E - E_Z).
\label{fimpurity}
\end{align}
The first term is comprised of two identical modes, each indicating the correlation between two conduction electrons with the same spin (up/up and down/down).
The latter two terms express the correlations between electrons with opposite spins (up/down and down/up).

Because of defect polarization in the magnetic field, the rates $\topp^{-1}$ and $\tsame^{-1}$ differ. The rate $\tsame^{-1}$, indicating the loss of coherence between two electrons which are identical (same spin), though passing through the graphene at different times, is given by\cite{vavilov}
\begin{equation}
\tsame^{-1}(\btotal) = \tinf^{-1} + [1-P(\btotal)^2]\tpol^{-1}.
\end{equation}
This may be interpreted as the decoherence rate proper (ie.~the dephasing by dynamic environment).
Here $\tinf^{-1}$ is the field-independent part of the decoherence rate (from e.g.~electron-electron interactions, non-magnetic dynamic defects) and $\tpol^{-1}$ is the rate of collision with polarizable magnetic defects. $P(\btotal)$ is the average polarization of the magnetic defects, due to the total field $\btotal$.

The correlation between conduction electrons of opposite spin (but equal kinetic energy) is dephased by a larger rate\cite{vavilov}
\begin{equation}
\topp^{-1}(\btotal) = \tinf^{-1} + [1+P(\btotal)^2]\tpol^{-1}.
\end{equation}
 Note that $\topp^{-1}$ is always dephased by magnetic impurities, even once they have been fully polarized ($P=1$); this extra dephasing is however not decoherence since it arises from static defects. Rather, the extra dephasing in $\topp^{-1}$ simply indicates that spin-up electrons consistently scatter differently than spin-down (this is analogous to the effect of spin-orbit interactions in UCF).

The Zeeman splitting of conduction electrons is also evident in Eq.~\eqref{fimpurity}: a shift of the opposite-spin UCF modes to energies $\delta E = \pm E_Z$, so these modes appear as side-peaks in the UCF energy correlation function.
In the theoretical literature these side-peaks are known as $m = \pm 1$ diffuson triplets\cite{vavilov,micklitz}. We have directly observed these side-peaks in graphene in Ref.~\onlinecite{nphys}, and the present device shows side-peaks as well, though weakly (Fig.~\ref{supp-ecorrelation}).

\begin{figure}
\includegraphics{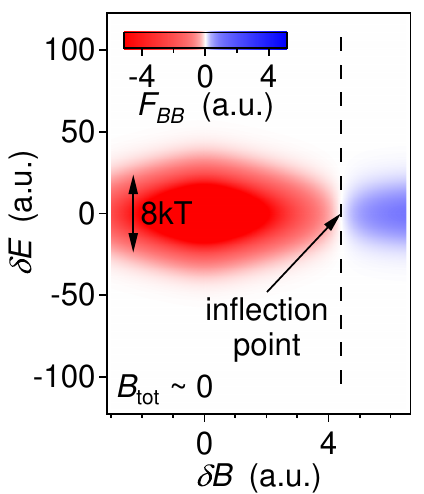}\includegraphics{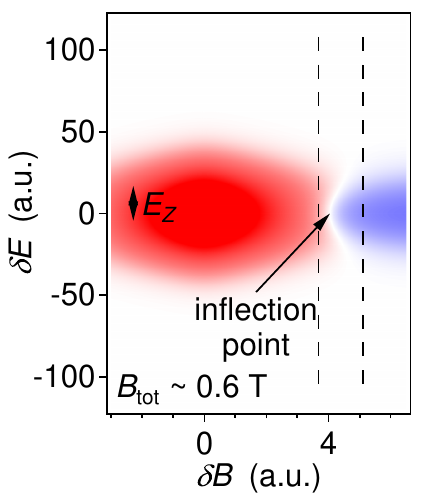}\\
\includegraphics{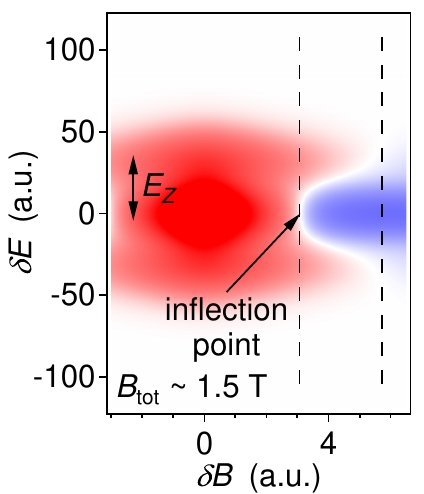}\includegraphics{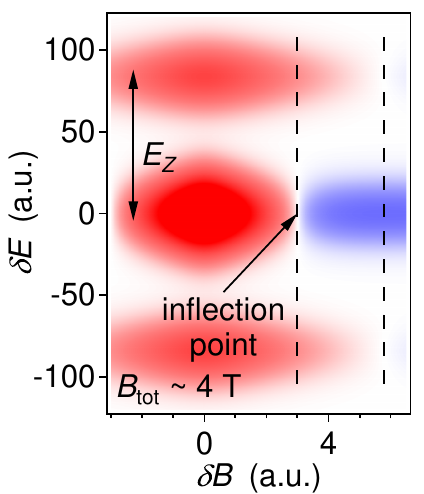}
\caption{\label{supp-smear}Second derivative $\frac{\partial^2}{\partial \delta B^2} F(\delta B, \delta E)$ of the simulated UCF correlation function \eqref{fimpurity}, for a case at 310 mK.
The vertical dashed lines indicate the fields $3\hbar\tsame^{-1}/2eD$ and $3\hbar\topp^{-1}/2eD$.
({\em Top-left}): Near-zero-field case, where $E_Z = 0$ and $P=0$.
({\em Top-right}): Intermediate field case where $E_Z = 2.6 kT$ and $P=71\%$.
({\em Bottom-left}): Intermediate field case where $E_Z = 6.5 kT$ and $P=98\%$.
({\em Bottom-right}): High field case, where $E_Z = 17 kT$ and $P=100\%$.
}
\end{figure}

Figure \ref{supp-smear} shows how Eq.~\eqref{fimpurity} evolves as the field is increased. Since it is not easy to see the inflection point in $F$ itself, we instead plot its second derivative in field, $\frac{\partial^2}{\partial \delta B^2} F(\delta B, \delta E)$. Again, we remind that $f(\delta B)$ only measures a cross section along $\delta E = 0$, thus the inflection point $\bip$ of $f(\delta B)$ corresponds to the point where $\frac{\partial^2}{\partial \delta B^2} F(\delta B, 0) = 0$.
We draw attention to a few important features of Fig.~\ref{supp-smear}:

\begin{figure}[t]
\includegraphics{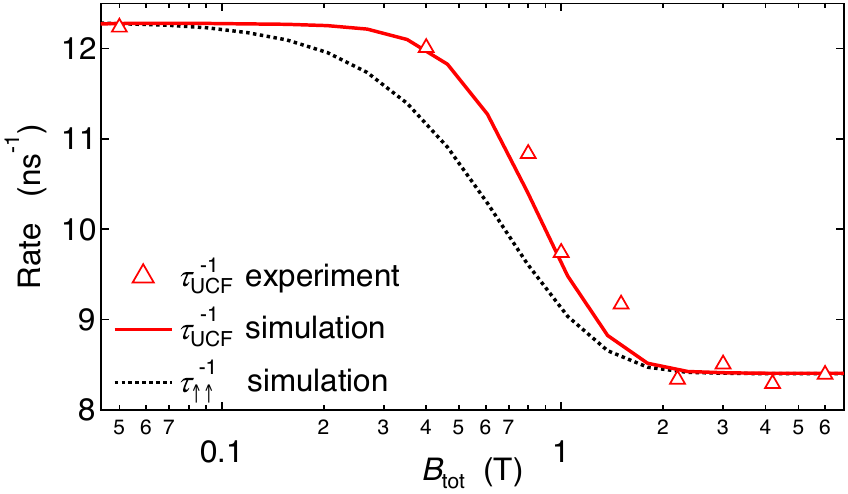}
\caption{\label{supp-bdep}Comparison between the rate $\tip^{-1}$ extracted from experimental 310 mK data (first cooldown) and from corresponding simulation with $g = 1.4$, compared to the decoherence rate $\tsame^{-1}$ from the same simulation. }
\end{figure}

\begin{itemize}
\item For very low fields we have $E_Z = 0$ and $P=0$ (thus $\topp = \tsame$). Here $F(\delta B,\delta E) = 4\tilde F[\tsame^{-1},T](\delta B,\delta E)$ has the the spinless correlation function's shape.
The inflection point here is (trivially) a good measure of decoherence, as $\tip^{-1} = \tsame^{-1} = \topp^{-1}$.
\item At intermediate fields where $E_Z \lesssim 8kT$, the side-peaks are close enough to influence the inflection point. This causes $\tip^{-1}$ to be influenced by both $\tsame^{-1}$ and $\topp^{-1}$ for these fields.
Figure~\ref{supp-bdep} shows precisely how $\tip^{-1}$ can deviate from the decoherence rate ($\tsame^{-1}$) over a range of intermediate fields.
\item When $E_Z$ is large enough ($E_Z \gtrsim 8kT$), the larger $\topp^{-1}$ does not matter as the side-peaks are too far away. The inflection point coincides with $3\hbar\tsame^{-1}/2eD$ once again, thus we measure the decoherence by $\tip^{-1}$.
\end{itemize}
The error $\tip^{-1}/\tsame^{-1}$ at intermediate fields is influenced by all parameters $T, E_Z, P, \tinf^{-1}, \tpol^{-1}$, so the only way to find its value is the direct numerical simulation as described above.

\subsubsection{Choice of polarization function}

We have chosen to use a polarization function of the form $P(\btotal) = \tanh(g \mu_B \btotal / 2 k_B T)$, which corresponds to the average magnetization of free spin-$\frac{1}{2}$ moments, normalized to $P(\infty) = 1$.
One might ask whether this is appropriate given that the model used above assumes classical magnetic defects, and is not necessarily applicable to spin-$\frac{1}{2}$ defects.\cite{vavilov}
We have compared the classical defect model to the much more involved quantum calculation\cite{vavilov} that includes the effects of inelastic scattering. At least for the central mode (a quantum calculation for the side-peaks is not available at this time\cite{vavilov}), the obtained $\tip^{-1}$ is nearly the same for the quantum and classical models, even for the spin $\frac{1}{2}$ case, provided that the $P = \tanh(g \mu_B \btotal / 2 k_B T)$ polarization function is used. (Such a correspondence does not extend to WL.)

\subsubsection{Parameters used in the primary text}

In order to generate the simulated curves of Fig.~3 in the primary text, we chose $\tmag^{-1} = 5~\mathrm{ns^{-1}}$ for all temperatures, based on the typical difference between low and high field. The value of $\tinf^{-1}$, which contains all non-magnetic decoherence mechanisms such as electron-electron interactions, is set by hand in each case to match the high-field $\tip^{-1}$.
The following table shows the values that were used in the simulations to produce the curves drawn in Fig.~3 of the primary text:

\begin{center}
\begin{tabular}{c|c|c}
	$T~~\mathrm{(K)}$ ~&~ $\tpol^{-1}~~\mathrm{(ns^{-1})}$ ~&~ $\tinf^{-1}~~\mathrm{(ns^{-1})}$ \\
	\hline
	 0.08 & 5 & 4.8 \\
	 0.2 & 5 & 6.4 \\
	 0.5 & 5 & 9.2 \\
	 1 & 5 & 14.5 \\
\end{tabular}
\end{center}

\end{document}